\documentclass[12pt]{article}
\usepackage{graphicx}

\newsavebox{\sboxpubnumber}
\newsavebox{\sboxpubdate}
\newcommand{\pubdate}[1]{\begin{lrbox}{\sboxpubdate}{#1}\end{lrbox}}
\newcommand{\pubnumber}[1]{\begin{lrbox}{\sboxpubnumber}{\begin{tabular}{l} #1 
\\
				 \usebox{\sboxpubdate}
				 \end{tabular}}
                           \end{lrbox}
                           \pubblock}
\newcommand{\Title}[1]{\begin{center} {\Large #1 } \end{center}}
\newcommand{\Author}[1]{\begin{center}{ \sc #1} \end{center}}
\newcommand{\Address}[1]{\begin{center}{ \it #1} \end{center}}

\newcommand{\pubblock}{\rightline{
			\usebox{\sboxpubnumber}}}
\newenvironment{Abstract}{\begin{quotation}  }{\end{quotation}}
\newenvironment{Presented}{\begin{quotation} \begin{center}
             PRESENTED AT\end{center}\bigskip
      \begin{center}\begin{large}}{\end{large}\end{center}
      \end{quotation}}
\newcommand{\Acknowledgements}{\bigskip  \bigskip \begin{center} \begin{large}
             \bf ACKNOWLEDGEMENTS \end{large}\end{center}}

\def\beq{\begin{equation}}
\def\eeq#1{\label{#1}\end{equation}}
\def\eeqn{\end{equation}}

\def\beqa{\begin{eqnarray}}
\def\eeqa#1{\label{#1}\end{eqnarray}}
\def\eeqan{\end{eqnarray}}

\let\bar=\overbar

\def\Dslash{\not{\hbox{\kern-4pt $D$}}}
\def\dslash{\not{\hbox{\kern-2pt $\del$}}}

\def\msb{{\bar{\ssstyle M \kern -1pt S}}}

\begin{document}

\begin{titlepage}
\pubdate{30th November 2001}                    
\pubnumber{~} 

\vfill
\Title{Inflationary Cosmology: Status and Prospects}
\vfill
\Author{Andrew R.~Liddle}
\Address{Astronomy Centre, University of Sussex, Falmer, Brighton BN1 
9QJ,~~~U.~K.}
\vfill
\begin{Abstract}
This article gives a brief overview of the status of attempts to constrain 
inflation using observations, and examines prospects for future developments. 
\end{Abstract}
\vfill
\begin{Presented}
    COSMO-01 \\
    Rovaniemi, Finland, \\
    August 29 -- September 4, 2001
\end{Presented}
\vfill
\end{titlepage}
\def\thefootnote{\fnsymbol{footnote}}
\setcounter{footnote}{0}

\section{Overview}

I begin with a brief reminder of what we are currently trying to achieve in 
cosmology. Personally, 
I'm interested in the following overall goals:
\begin{itemize}
\item To obtain a physical description of the Universe, including its global 
dynamics and matter content.
\item To measure the cosmological parameters describing the Universe, and to 
develop a fundamental understanding of as many of those parameters as possible.
\item To understand the origin and evolution of cosmic structures.
\item To understand the physical processes which took place during the extreme 
heat and density of the early Universe.
\end{itemize}
Over recent years, much progress has been made on all of these topics, to the 
extent that it is widely believed amongst cosmologists that we may stand on the 
threshold of the first precision cosmology, in which the parameters necessary to 
describe our Universe have been identified and, in most cases at least, measured 
to a satisfying degree of precision. Whether this optimism has any grounding in 
reality remains to be seen, though so far the signs are promising in that the 
basic picture of cosmology, centred around the Hot Big Bang, has time and again 
proven the best framework for interpretting the constantly improving 
observational situation.

In particular, the process of cosmological parameter estimation is well 
underway, thanks to observations of distant Type Ia supernovae, of galaxy 
clustering, and of
the cosmic microwave background. These have established a standard cosmological 
model, where the Universe is dominated by dark energy, contains substantial dark 
matter, and with the baryons from which we are made comprising only around 5\%. 
Overall this model can be described by around ten parameters (e.g.~see Wang et 
al.~\cite{WTZ}), and the viable region of parameter space is starting to shrink 
under pressure from observations. 
However, it is worth bearing in mind that we seek high precision determinations 
at least in part because they ought to shed light on fundamental physics, and 
there progress has been less rapid. Some parameters are likely to have no 
particular fundamental importance (for instance, there would probably be little 
fundamental significance were the Hubble constant to turn out to be $63 \, {\rm 
km \, s}^{-1} \, {\rm Mpc}^{-1}$ rather than say $72 \, {\rm km \, s}^{-1} \, 
{\rm Mpc}^{-1}$, though accurate measurements of the Hubble constant are 
essential in determining other parameters), but the 10\% or so measured accuracy 
of the baryon density is 
to be set against the lack of even an order-of-magnitude theoretical 
understanding thus far.

\section{Inflationary cosmology}

This article focusses on the last two of the goals listed at the start of the 
previous 
section. The claim is that during the very early Universe, a physical process 
known as {\bf inflation} took place, which still manifests itself in our present 
Universe via the perturbations it left behind which later led to the development 
of structure in the Universe. By studying those structures, we hope to shed 
light on whether inflation occurred, and by what physical mechanism.

I begin by defining inflation. The scale factor of the Universe at a given time 
is measured by the scale factor $a(t)$. In general a homogeneous and isotropic 
Universe has two characteristic length scales, the curvature scale and the 
Hubble 
length. The Hubble length is more important, and is given by
\begin{equation}
cH^{-1} \quad {\rm where} \quad H \equiv \frac{\dot{a}}{a} \,.
\end{equation}
Typically, the important thing is how the Hubble length is changing with time as 
compared to the expansion of the Universe, i.e.~what is the behaviour of the 
{\bf comoving Hubble length} $H^{-1}/a$?

During any standard evolution of the Universe, such as matter or radiation 
domination, the comoving Hubble length increases. It is then a good estimate of 
the size of the observable Universe. {\bf Inflation} is defined as any epoch of 
the Universe's evolution during which the comoving Hubble length is decreasing
\begin{equation}
\frac{d\left(H^{-1}/a\right)}{dt} < 0 \Longleftrightarrow \ddot{a}>0 \,,
\end{equation}
and so inflation corresponds to any epoch during which the Universe has 
accelerated expansion. During this time, the expansion of the Universe outpaces 
the growth of the Hubble radius, so that physical conditions can become 
correlated on scales much larger than the Hubble radius, as required to solve 
the horizon and flatness problems.

As it happens, there is very good evidence from observations of Type Ia 
supernovae that the Universe is {\em presently} accelerating \cite{Sn}. This is 
usually 
attributed to the presence of a cosmological constant. This is clearly at some 
level good news for those interested in the possibility of inflation in the 
early Universe, as it indicates that inflation is possible in principle, and 
certainly that any purely theoretical arguments which suggest inflation is not 
possible should be treated with some skepticism.

If the Universe contains a fluid, or combination of fluids, with energy density 
$\rho$ and pressure $p$, then
\begin{equation}
\ddot{a}>0 \Longleftrightarrow \rho + 3p < 0 \,,
\end{equation}
(where the speed of light $c$ has been set to one). As we always assume a 
positive energy density, inflation can only take place if the Universe is 
dominated by a material which can have a negative pressure. Such a material is a 
scalar field, usually denoted $\phi$. A homogeneous scalar field has a kinetic 
energy and a potential energy $V(\phi)$, and has an effective energy density and 
pressure given by
\begin{equation}
\rho = \frac{1}{2} \dot{\phi}^2 + V(\phi) \quad ; \quad p = \frac{1}{2} 
\dot{\phi}^2 - V(\phi) \,.
\end{equation}
The condition for inflation can be satisfied if the potential dominates.

A model of inflation typically amounts to choosing a form for the potential, 
perhaps supplemented with a mechanism for bringing inflation to an end, and 
perhaps may involve more than one scalar field. In an ideal world the potential 
would 
be predicted from fundamental particle physics, but unfortunately there are many 
proposals for possible forms. Instead, it has become customary to assume that 
the potential can be freely chosen, and to seek to constrain it with 
observations. A suitable potential needs a flat region where the potential can 
dominate the kinetic energy, and there should be a minimum with zero potential 
energy in which inflation can end. Simple examples include $V = m^2 \phi^2/2$ 
and $V = \lambda \phi^4$, corresponding to a massive field and to a 
self-interacting field respectively. Modern model building can get quite 
complicated --- see Ref.~\cite{LR} for a review.

By far the most important aspect of inflation is that it provides a possible 
explanation for the origin of cosmic structures. The mechanism is fundamentally 
quantum mechanical; although inflation is doing its best to make the Universe 
homogeneous, it cannot defeat the uncertainty principle which ensures that 
residual inhomogeneities are left over.\footnote{For a detailed account of the 
inflationary model of the origin of structure, see Ref.~\cite{LL}.} These are 
stretched to astrophysical scales by the inflationary expansion. Further, 
because these are determined by fundamental physics, their magnitude can be 
predicted independently of the initial state of the Universe before inflation. 
However, the magnitude does depend on the model of inflation; different 
potentials predict different cosmic structures.
Inflation models generically predict two independent types of perturbation,
density perturbations (scalar perturbations) and gravitational waves (tensor 
perturbations).

\section{The key tests of inflation}

In the remainder of this article, I will be interested solely in inflation as a 
model for the origin of structure; while it serves a useful purpose in solving 
the horizon and flatness problems these are no longer likely to provide further 
tests of the model. Indeed, the aim is to consider inflation as the {\em sole} 
origin of structure, since it is impossible to exclude an admixture of 
inflationary perturbations at some level even if another mechanism becomes 
favoured.

The key tests of inflation can be summarized in one very useful sentence, which 
lists in bold-face six key predictions that we would like to 
test. 
\begin{quote}
The {\em simplest} models of inflation predict {\bf nearly power-law} spectra of 
{\bf adiabatic}, {\bf gaussian} scalar and {\bf tensor} perturbations in their 
{\bf growing mode} in a {\bf spatially-flat} Universe.
\end{quote}
However some tests are more powerful than others, because some are 
predictions only of the simplest inflationary models. In what follows, it will 
be important to distinguish between tests of the inflationary idea itself, 
versus tests of different inflationary models or classes of models.

Before progressing to a discussion of the status of these tests, it is useful to 
define some terminology fairly precisely. In this article, a useful {\bf test} 
of 
a model is one which, if failed, leads to rejection of that model. The concept 
of a test is to be contrasted with {\bf supporting evidence}, which is 
verification of a prediction which, while not generic, is seen as indicative 
that the model is correct. A model can also accrue supporting evidence via its 
rivals failing to survive tests that they are put to.

\subsection{Spatial flatness}

All the standard models of inflation give a flat Universe, and this used to be 
advertised as a robust prediction. Unfortunately we now realise that it is 
possible to make more complicated models which can give an open Universe 
\cite{open}. 
Spatial geometry therefore does not constitute a test of the inflationary 
paradigm, as if the Universe were not flat there would remain viable 
inflationary models. However the recent microwave anisotropy experiments showing 
good consistency with spatial flatness provide good supporting evidence for the 
simplest inflation 
models.

\subsection{Gaussianity and adiabaticity}

If inflation is driven by more than one scalar field, there is a possibility of 
having isocurvature perturbations as well as adiabatic ones. The mechanism is 
that during inflation the other fields also receive perturbations. If they 
survive to the present (in particular, if they become the dark matter), this 
will give an isocurvature perturbation. As far as model building is concerned, 
such isocurvature perturbations could dominate, though this is now excluded by 
observations. More pertinently, the observed structures could be due to an 
admixture of adiabatic and isocurvature perturbations, and indeed those modes 
could be 
correlated.

Such models would also give either gaussian or nongaussian perturbations, and
nongaussian adiabatic models are also possible.  Whether observed nongaussianity
rules out inflation depends very much on the type of nongaussianity observed;
for instance chi-squared distributed fluctuations could easily be produced if
the leading contribution to the perturbations is quadratic in the scalar field
perturbation, while if any coherent spatial structures were seen, such as line
discontinuities in the microwave background, it would be futile to try and
produce them using inflation.

\begin{figure}[htb]
    \centering
    \includegraphics[height=9cm]{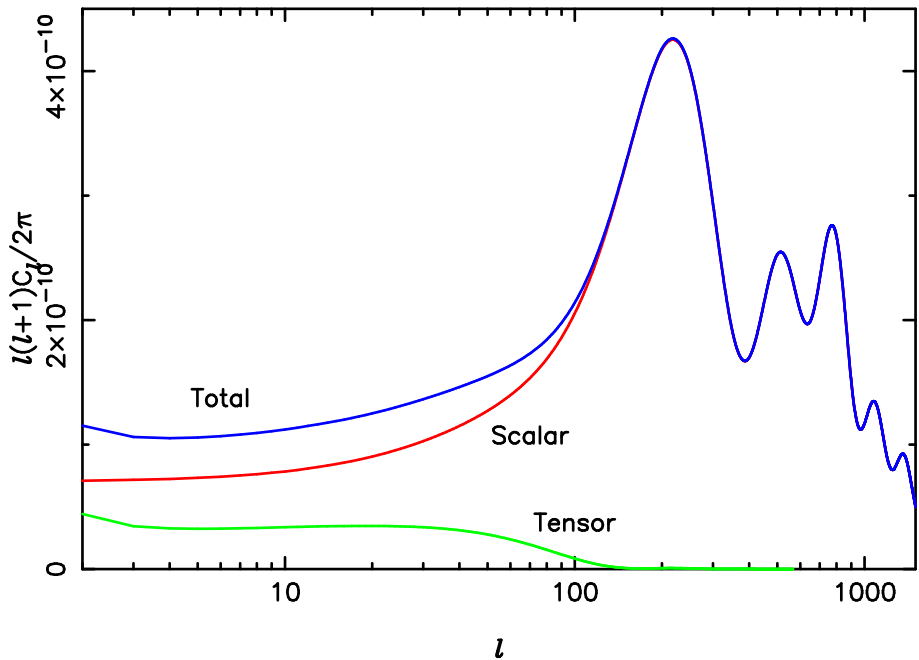}
    \caption{Scalars and tensors give different contributions 
to the anisotropies. While only the total spectrum is observable, provided the 
spectra have sufficiently simple forms this can be decomposed into the two 
parts, the tensors giving excess perturbations at small $\ell$.}
    \label{f:cl_st}
\end{figure}

\subsection{Vector and tensor perturbations}

All inflation models produce gravitational waves at some level, and if seen they 
can provide extremely strong supporting evidence for inflation. They are not 
however a test, as their absence could mean an inflation model where their 
amplitude is undetectably small. The best way to detect them is by their 
contribution to large-angle anisotropies, as shown in Figure~\ref{f:cl_st}.

By contrast, known inflation models do not produce vector perturbations, and 
indeed inflation will destroy any pre-existing ones. If detected, at the very 
least they would present a challenge for inflation model builders. It would be 
interesting to make a comprehensive study to confirm whether detection of vector 
modes would be sufficient to exclude inflation as the sole origin of structure.

\subsection{Growing mode perturbations}

A key property of inflationary perturbations is that they were created in the 
early 
Universe and evolved freely from then. Although a general solution to the 
perturbation equations has two modes, growing and decaying, only the growing 
mode will remain by the time the perturbation enters the horizon. This leads 
directly to the prediction of an oscillatory structure in the microwave 
anisotropy power spectrum \cite{peaks}. The 
existence of such 
a structure is a robust prediction of inflation; if it is not seen then 
inflation cannot be the sole origin of structure.

The most significant recent development in observations pertaining to inflation 
is the first clear evidence for multiple peaks in the spectrum, seen by the DASI 
\cite{DASI} and Boomerang \cite{Boom01} experiments. This is a crucial 
qualitative test which inflation appears to have passed, and which could have 
instead provided evidence against the entire inflationary paradigm. These 
observations lend great support to inflation, though it must be stressed that 
they are not able to `prove' inflation, as it may be that there are other ways 
to produce such an oscillatory structure \cite{Neil}.

\section{Present and future}

\subsection{The current status of inflation}

The best available constraints come from combining data from different sources; 
for two recent attempts see Wang et al.~\cite{WTZ} and Efstathiou et 
al.~\cite{Eetal}. Suitable data include observations of the recent dynamics of 
the Universe using Type Ia supernovae, cosmic microwave anisotropy data, and 
galaxy correlation function data.

Currently inflation is a massive qualitative success, with striking agreement 
between the predictions of the simplest inflation models and observations. In 
particular, the locations of the microwave anisotropy power spectrum peaks are 
most simply interpretted as being due to an adiabatic initial perturbation 
spectrum in a spatially-flat Universe. The multiple peak structure strongly 
suggests that the perturbations already existed at a time when their 
corresponding scale was well outside the Hubble radius. No unambiguous evidence 
of nongaussianity has been seen.

Quantitatively, however, things have some way to go. At present the best that 
has been done is to try and constrain the parameters of the power-law 
approximation to the inflationary spectra. The gravitational waves have not been 
detected and so their amplitude has only an upper limit and their spectral index 
is not constrained at all. The current situation can be summarized as follows.
\begin{description}
\item[Amplitude $\delta_{{\rm H}}$:] COBE determines this (assuming no 
gravitational waves) to about ten percent accuracy (at one-sigma) as 
approximately
$\delta_{{\rm H}} = 1.9 \times 10^{-5} \, \Omega_0^{-0.8}$ on a scale close to 
the present Hubble 
radius (see Refs.~\cite{BLW,LL} for accurate formulae).
\item[Spectral index $n$:] This is thought to lie in the range $0.8 < n < 1.05$ 
(at 95\% confidence). It would be extremely interesting were the 
scale-invariant case, $n=1$, to be convincingly excluded, as that would be clear 
evidence of dynamical processes at work, rather than symmetries, in creating the 
perturbations.
\item[Gravitational waves $r$:] Measured in terms of the relative contribution 
to large-angle microwave anisotropies, the tensors are currently constrained to 
be no more than about 30\%.
\end{description}

\subsection{Prospects for the future}

It remains possible that future observations will slap us in the face and lead 
to inflation being thrown out. But if not, we can expect an incremental 
succession of better and better observations, culminating (in terms of 
currently-funded projects) with the {\em Planck} satellite \cite{Planck}. Faced 
with observational data of exquisite quality, an initial goal will be to test 
whether the simplest models of inflation continue to fit the data, meaning 
models with a single scalar field rolling slowly in a potential $V(\phi)$ which 
is then to be constrained by observations. If this class of models does remain 
viable, we can move on to reconstruction of the inflaton potential from the 
data.

{\em Planck}, currently scheduled for launch in February 2007, should be highly 
accurate. In particular, it should be able to measure the spectral index to an 
accuracy better than $\pm 0.01$, and detect gravitational waves even if they are 
as little as 10\% of the anisotropy signal. In combination with other 
observations, these limits could be expected to tighten significantly further, 
especially the tensor amplitude. Such observations would rule out almost all 
currently known inflationary models. Even so, there will be considerable 
uncertainties, so it is important not to overstate what can be achieved.

\begin{figure}[p]
    \centering
    \includegraphics[height=9cm]{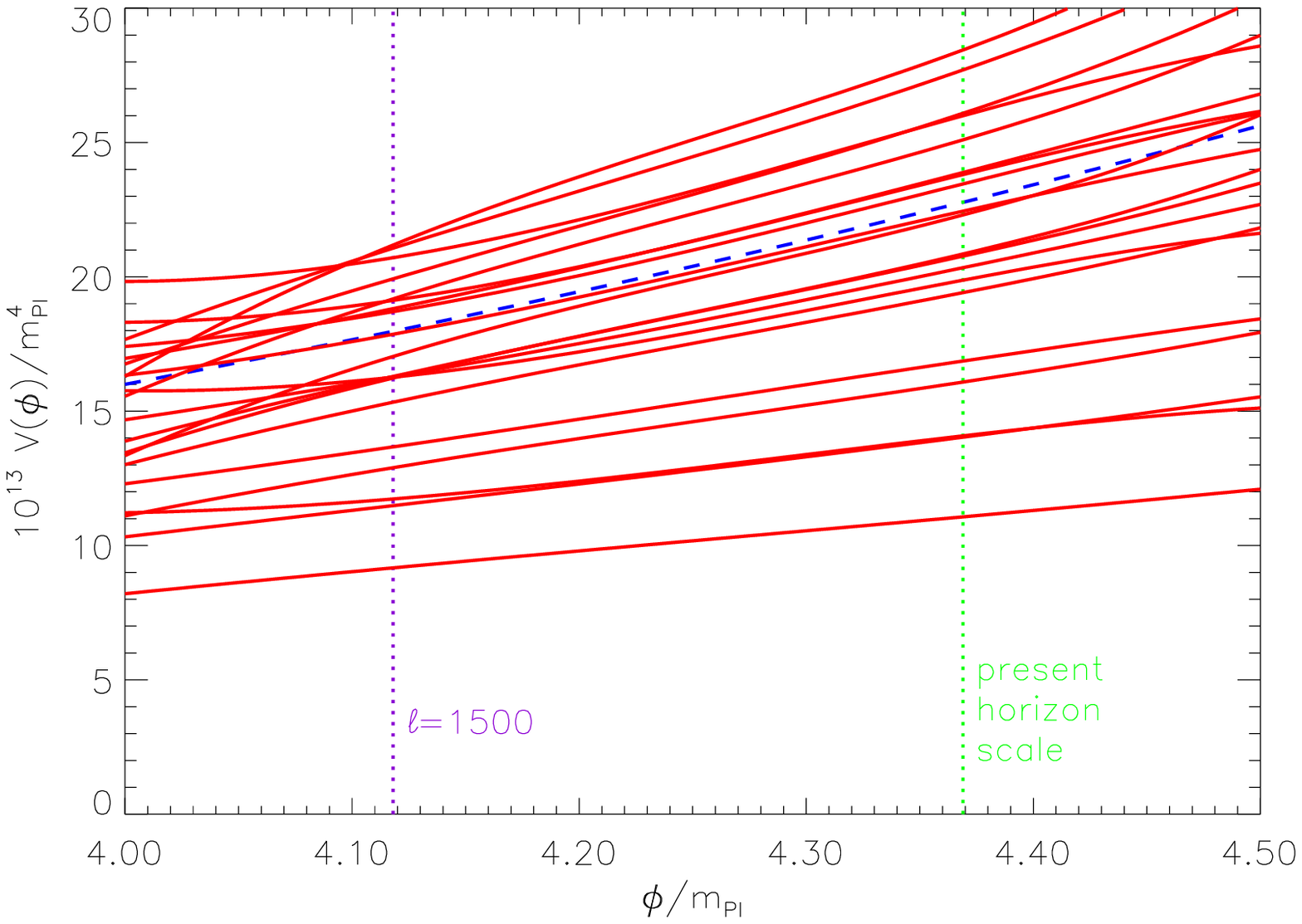}
    \includegraphics[height=9cm]{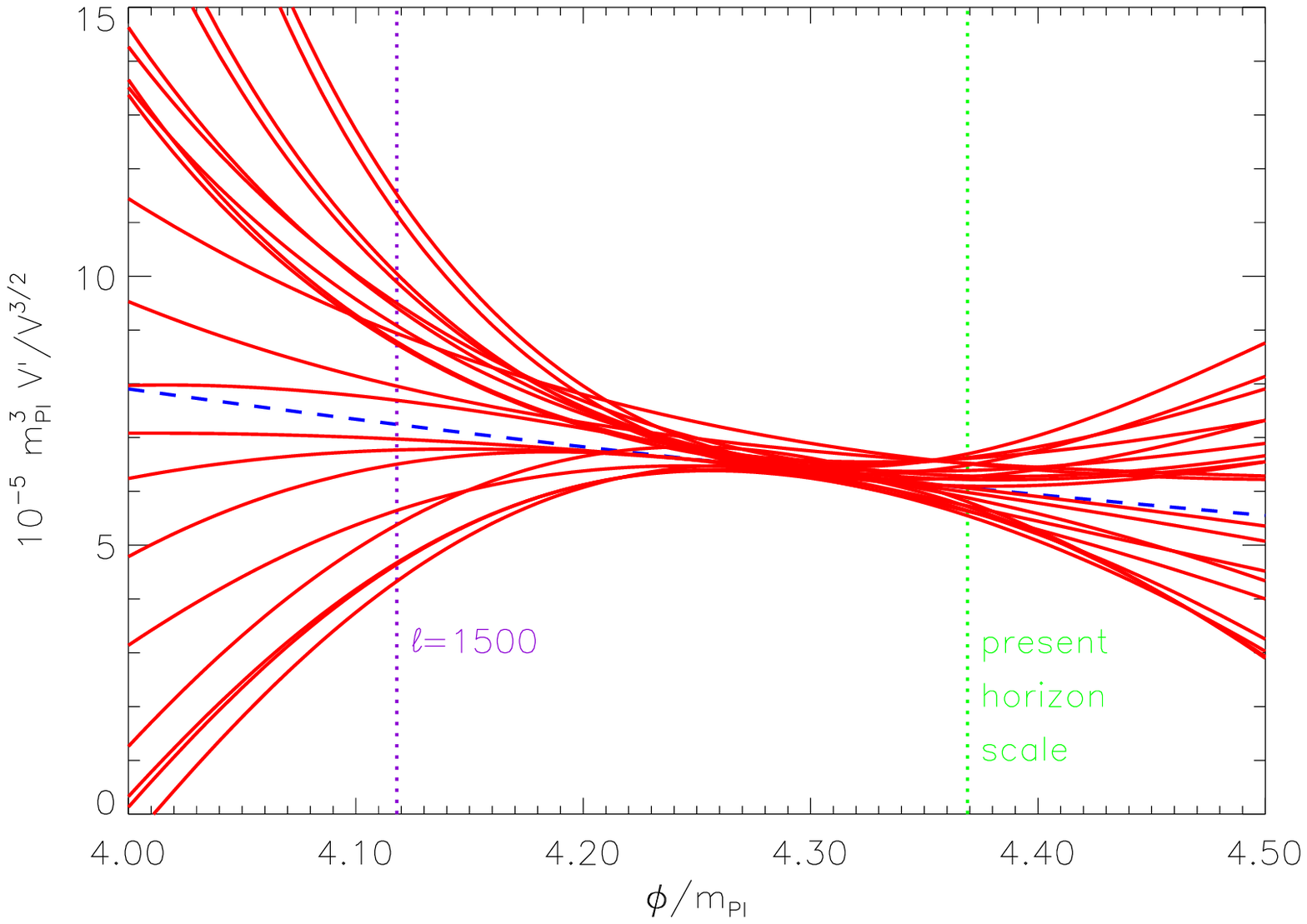}
    \caption{Sample reconstruction of a potential, where the 
dashed line shows the true potential and the solid lines are thirty Monte Carlo 
reconstructions (real life can only provide one). The upper panel shows the 
potential itself which is poorly determined. However some combinations, such as 
$(dV/d\phi)/V^{3/2}$ shown in the lower panel, can be determined at an accuracy 
of a few percent. See Ref.~\cite{GL} for details.}
    \label{f:recon}
\end{figure}

Reconstruction can only probe the small part of the potential where the field 
rolled while generating perturbations on observable scales. We know enough about 
the configuration of the {\em Planck} satellite to be able to estimate how well 
it should perform. Ian Grivell and I recently described a numerical technique 
which gives an optimal construction \cite{GL}. Results of an example 
reconstruction are shown in Figure~\ref{f:recon}, where it was assumed that the 
true potential was $\lambda \phi^4$. The potential itself is not well determined 
here (the tensors are only marginally detectable), but certain combinations, 
such as $(dV/d\phi)/V^{3/2}$, are 
accurately constrained and would lead to high-precision constraints on inflation 
model parameters.

\section{New directions}

While the simplest models of inflation provide an appealing simple framework 
giving excellent agreement with observations, it is important to consider 
whether a similar outcome might arise from a more complicated set-up that has 
better motivation from fundamental physics. There are some new ideas circulating 
in this regard, of which I will highlight just two related ones.

\subsection{The braneworld}

Particle physicists generally tend to believe that our Universe really possesses 
more than three spatial dimensions. Previously it has been assumed that the 
extra ones were ``curled up'' to be unobservably small. A new idea is the {\bf 
braneworld}, which proposes that at least one of these extra dimensions might be 
relatively large, with us constrained to live on a three-dimensional {\bf brane} 
running 
through the higher-dimensional space. Gravity is able to propagate in the full 
higher-dimensional space, which is known as the {\bf bulk}.

This radical idea has many implication for cosmology, both in the present and 
early Universe, and so far we have only scratched the surface of possible new 
phenomena. Already many exciting results have been obtained, and I'll just 
consider a few pertinent questions.

~

\noindent
{\em 1.~Are there modifications to the evolution of the homogeneous Universe?}\\
The answer appears to be yes; for example in a simple scenario (known as 
Randall--Sundrum Type II \cite{RSII}) the Friedmann equation is modified at high 
energies so that, after some simplifying assumptions, it reads \cite{bin}
\begin{equation}
H^2 = \frac{8\pi G}{3} \left( \rho+ \frac{\rho^2}{2\lambda} \right) \,,
\end{equation} 
where $\lambda$ is the tension of the brane. This recovers the usual cosmology 
at low energies $\rho \ll \lambda$, but otherwise we have new behaviour. This 
opens new opportunities for model building, see for example Ref.~\cite{CLL}.

~

\noindent
{\em 2.~Are inflationary perturbations different?}\\
Again the answer is yes --- there are modifications to the formulae giving 
scalar and tensor perturbations \cite{braneperts}. Unfortunately the main effect 
of this is to introduce new degeneracies in interpretting observations, as a 
potential can always be found matching observations for any value of $\lambda$ 
\cite{LT}. The initial perturbations therefore cannot be used to test the 
braneworld scenario.

~

\noindent
{\em 3.~Do perturbations evolve differently after they are laid down on large 
scales?}\\
The answer here is less clear. It is certainly possible that perturbation 
evolution is modified even at late times. For example perturbations in the bulk 
could influence the brane in a way that couldn't be predicted from brane 
variables alone. Whether there is a significant effect is unclear and is likely 
to be model dependent.

\subsection{The Ekpyrotic Universe}

It has recently been proposed that the Big Bang is actually the result of the
collision of two branes, dubbed the Ekpyrotic Universe \cite{ekpyrotic}; this
scenario is discussed in detail elsewhere in these proceedings.  It has
been claimed that this scenario can provide a resolution to the horizon and
flatness problems, essentially because causality arises from the
higher-dimensional theory and allows a simultaneous Big Bang everywhere on our
brane, though existing implementations solve the problem by hand in the initial
conditions.  As I write this, it remains unclear how to successfully describe
the instant of collision between the two branes (the singularity problem), and
considerable controversy surrounds whether or not the scenario can also generate
nearly scale-invariant adiabatic perturbations \cite{ekperts}.  Both aspects are
required to make it a serious rival to inflation.

\section{Summary}

These are extremely good times for the inflationary cosmology. Its qualitative 
predictions have time and again provided the simplest interpretation of 
observational data, while historical rivals such as cosmic strings 
\cite{Strings} have faded away.
There is every prospect that upcoming observations will provide high-accuracy 
constraints on the models devised. The necessary theoretical ingredients all 
appear to be in place to allow predictions of the required sophistication to be 
made. We keenly await new observational data.

\Acknowledgements

The author was supported in part by the Leverhulme Trust.



\end{document}